\documentstyle [aps,epsfig,amsfonts]{revtex}

\def\(({\left(}
\def\)){\right)}
\def\[[{\left[}
\def\]]{\right]}

\def \q {q_{12}}
\def \l {\langle}

\newcommand \be{\begin{equation}}

\def \bi{\bibitem}

 \def\(({\left(}
 \def\)){\right)}
\def\bi{\bibitem}

\def \ov{\over}

\def \a{\alpha}
\def \b{\beta}

\def \e{{\rm e}}

\def \beqna{\begin{eqnarray}}
\def \eeqna{\end{eqnarray}}
\def \beq{\begin{equation}}
\def \eeq{\end{equation}}
\def \ln{{\rm ln}}
\def \ov{\over}

\def \ol{\overline}
\def \a{\alpha}

\def \b{\beta}
\def \ri{\right}
\def \sig{{\bf\sigma}}
\def \l{\left}
\def \ln{{\rm ln}}
\def \ra{\rangle}
\def \ab2{\alpha\beta^2}
\def \ra{\rangle}

\newcommand {\Ga} {\Gamma}
\newcommand {\si} {\sigma}
\newcommand \bea {\begin{eqnarray} \nonumber }
\newcommand \ee {\end{equation}}
\newcommand \eea {\end{eqnarray}}
 \newcommand \eps {\epsilon}

\newcommand \la {\langle}

\begin{document}
\title{Temperature evolution and bifurcations
of metastable states in mean-field spin glasses, with
connections with structural glasses}
\author{ Alain Barrat(*), Silvio Franz(*) and Giorgio Parisi(**)}
\address{
(*) International Center for Theoretical Physics\\
Strada Costiera 11,
P.O. Box 563,
34100 Trieste (Italy)\\
(**) Universit\`a di Roma ``La Sapienza''\\
Piazzale A. Moro 2, 00185 Rome (Italy)\\
e-mail: {\it barrat@ictp.trieste.it,
 franz@ictp.trieste.it, parisi@roma1.infn.it}}
\date{February 1997}
\maketitle

\begin{abstract}
The correlations of the free-energy landscape of mean-field spin glasses
at different temperatures are investigated, concentrating on models with a
first order freezing transition. Using a ``potential function''
we follow the metastable states of the model in temperature, and discuss
the possibility of level crossing (which we do not find) and multifurcation
(which we find). The dynamics at a given temperature
starting from an equilibrium configuration at a different temperature
is also discussed. In presence of multifurcation, we find that the
equilibrium is never achieved, leading to aging behaviour at
slower energy levels than usual aging.
The relevance of the observed mechanisms for real structural glasses
is discussed, and some numerical simulations of a soft sphere
model of glass are presented.
\end{abstract}

\section{Introduction}

Free-energy landscapes in high dimensional
spaces  have longly been used as metaphores for describing the
physics of complex systems as glasses and spin-glasses,
and also  proteins and evolutionary fitness landscapes \cite{free-en-lan}.
The basic idea of this approach is that complex systems
dynamics can be viewed as a search for optima in a rough hypersurface.
Although free-energy surfaces can in principle be defined for
large classes of finite dimensional models, the actual construction
of such functions has been achieved only in the case of long-range
disordered system (mean-field spin glasses), which relevance
for finite dimensional spin glasses has been subject of a long debate,
growing evidence \cite{sim-num} pointing in the direction that
mean-field theory is a good starting point to describe finite dimensional
physics. Thouless Anderson
and Palmer \cite{TAP} (TAP) showed that stable and metastable states of
long-range spin glasses are associated to minima of a suitable
free-energy, which is a random function in a $N$ ($\to\infty$)
dimensional space. The complex phenomenology of equilibrium
spin-glasses can be read as a set of propositions about the low
lying minima of the TAP free-energy, and the barriers separating
them.

Due to the random character of the TAP free-energy function,
analytic statements on the structure of the stationary point
have forcedly statistical character. Various techniques
have been invented to study the minima
of TAP free-energy, and the structure of the stable
and metastable minima for fixed external parameter is known in
great detail, and gives a coherent picture of the glassy transition.
Two classes of models are known, according to the order of the freezing
transition. In models like the Sherrington-Kirkpatrick models,
which display a second order phase transition, the metastable states
of the TAP free-energy do not play an important physical role.
A second order transition suggests a mechanism of bifurcation
(or rather multifurcation) of the paramagnetic minimum as $T_c$ is crossed.
In off-equilibrium dynamics it is found that all the extensive quantities
tend to their equilibrium values for large times.
Whereas these mean-field models, with continuous transition, seem
to apply for the description of real spin glasses, a second class
of models, like e.g. the Potts glass,
 show a first order freezing transition, and seem closer to describing
the physics of structural glasses. Indeed, for these models occurs
a purely dynamical transition where the relaxation time
diverges, while the static thermodynamic quantities show
singularities only at a lower temperature. Below the
dynamical transition temperature, metastable
states dominate the physics, and,
dynamically, the extensive quantities do not tend to their
equilibrium values if a random initial condition is chosen.
Statically, the partition function is dominated by
metastable states (between the statical and the dynamical transitions,
by an exponentially large number of mutually inaccessible states)
that the system is unable to reach dynamically.

The scenario in which the barriers between metastable states
are infinite, and where a quenched system never reaches any
of these states, is clearly linked to the mean-field approximation.
For finite systems, metastable states have a finite lifetime,
and the system should be able to find them in a finite time. This
time, and the states the system is able to find,
can depend for example on the cooling rate.
A modified scenario
would include ``activated processes'' and suppress the divergence
of relaxation times at the dynamical transition, replacing it by
a rapid increase (with divergence only at the static transition).

This picture, as was already advocated in
\cite{kithirum,ktglass,kirkwol} and more recently in
\cite{parisi_verre},
could be relevant for real glasses: indeed, the glass
transition temperature is also a purely
dynamical quantity, defined by the fact that the relaxation time reaches
a certain value, the existence of a static transition at a lower
temperature still being subject of debate. Below this
temperature, the system remains out of equilibrium
for all available time scales, and static quantities are not
reached. In the same way, the aforementioned scenario
would yield a glass transition (corresponding to a large but finite
value of the relaxation time) occuring above the static transition,
and a dynamical evolution resulting from a mixture of
mean-field like dynamics and activated processes.

In this context, the relevance of mean-field studies depends
on the various time scales involved: if the barriers between
metastable states are low, activated processes are fast and will
dominate the evolution; if on the contrary the energy barriers
are finite but large, there will exist time windows in which
the mean-field scenario will hold.

To address this question, we have therefore to
gain knowledge on the metastable states, both
statically and dynamically, for the mean-field models, and to compare the
emerging picture and dynamical scenarios with the real world,
or at least with numerical simulations.

If the structure of the TAP minima for fixed temperature is
rather well known \cite{mpv,kurparvir},
a much less coherent picture is available for
the correlation of the free-energy landscapes for different temperatures.
In the Sherrington-Kirkpatrick model, the study of \cite{kondor,franey}
have revealed ``chaotic temperature dependence'' of the low lying states.
States of equilibrium at different temperatures are, no matter how small
the temperature differences, as uncorrelated as they can.
In some other models the chaotic property is absent. A clear
example is the spherical $p$-spin model, where the homogeneity
of the Hamiltonian implies that the order of the free-energy minima does not
depend on temperature, so that, in the whole low-temperature phase,
the statics is given by the same low-lying states.
Some general conclusions  about the fate of TAP minima under temperature
changes can be drawn on the basis of smoothness of the TAP free-energy
as a function of temperature. For example, absolutely stable minima
can not disappear or multifurcate for an infinitesimal change of the
temperature, and the generation of new stationary points has to pass
by a marginally stable situation.

Two issues appear to be relevant for the description of correlations
of the landscapes for different temperatures: level bifurcations
and level crossing.
Therefore, in this paper we try to gain some generic
insights on these topics, by
addressing the issue of following the TAP states in temperature
for a  spherical model which displays first order glassy transition.
Differently from the $p$-spin model, the Hamiltonian is not homogeneous.
We expect therefore that its behavior is generic in the class of models
with first order transition. The analysis is performed
with the aid of a recent method where the metastable
states are associated to local minima of some macroscopic ``potential''
function of the spin-glass order parameter.
The basic idea of this approach is that the free-energy manifold
can be probed introducing an external field pointing in the direction
of some typical equilibrium configuration \cite{I,remi,fp97}.
In section two, we review the construction of the potential of ref. \cite{I}
and extend the discussion to some properties not mentioned there.
In section three we use this potential to follow metastable states in
temperature, and discuss the possibility of multifurcation.

In section four,
we use another powerful approach, a dynamical one, with appropriate
initial conditions: the dynamics of a system thermalized at a certain
temperature, and then brought at another, also allows to explore
the phase space of the system \cite{I,babumez}.
We show that the two methods
yield the same results, and use
moreover this dynamical study to tackle
another relevant issue: the dynamical behavior
of the  systems when a TAP solution bifurcates. In particular
the problem whether the system
will fall into one of the new valleys
or will be unable to decide where to fall and
age forever.

After having described these mechanisms for the considered
mean-field models, we tentatively compare them in section five
to the case of
real glasses, via numerical simulations of a soft sphere glass.
In particular, the dynamical mean-field approach of thermalized initial
conditions can be thought of as a previous very slow cooling
to a certain temperature, followed by a rapid change. The
study of the energy reached with various cooling rates in the
simulated system shows the relevance of the mean-field
scenario, in the available time window.

\section{Studied models; the potential; previous results}

The class of models we consider is defined by $N$ (real) spins
${\bf{s}}=\{s_1,\cdots,s_N\}$ interacting through
a  Hamiltonian $H({\bf{s}})$
and a global (spherical) constraint $\sum_i s_i^2 = N$. The
Hamiltonian is random, Gaussian, with correlations
\be
\overline{ H({\bf{s}})H({\bf{s'}})} = N f(q_{\bf{s s'}}),
\ee
where $q_{\bf{s s'}}=1/N \sum_i s_i s'_i$ is  the overlap between the
configurations ${\bf{s}}$ and ${\bf{s'}}$. If
$f$ is a polynomial function, the Hamiltonian can be presented
as a linear combination of terms of the type
\be
H_p({\bf{s}})= - \sum_{1 \leq i_1 < i_2 \cdots < i_p \leq N}
J_{i_1 i_2 \cdots i_p} s_{i_1}s_{i_2}\cdots s_{i_p},
\end{equation}
with Gaussian independent   couplings $J_{i_1 i_2 \cdots i_p}$,
 with zero mean and variance
$p!/(2N^{p-1})$. It is easy to see that
$\overline{ H_p({\bf{s}})H_{p'}({\bf{s'}})}=\delta_{p,p'} q_{\bf{s s'}}^p /2$.
As we will see in the following, the purely monomial case, the
so called $p$-spin model, has remarkably simple properties under temperature
changes \cite{kurparvir}
thanks to the homogeneity of the Hamiltonian under contemporary
rescaling of all the variables. In order to study the generic behavior
it is therefore necessary to consider inhomogeneous Hamiltonian,
giving rise to  non monomial correlation functions.
The specific form of the function $f(q)$ we will
use  in our examples will be mainly $f(q)=1/2(q^3+q^4)$,
however the results will be generic for inhomogeneous Hamiltonian
verifying the condition that $f''(q)^{-3/2}f'''(q)$ is  monotonically
decreasing with $q$ for all $q$ (this ensures that the transition
is a discontinuous one).

In that case the statics of the model is described by
\begin{itemize}
\item a high temperature phase, for $T > T_d$, where the dominant
contribution is given by a paramagnetic state;
\item a temperature range $T_d > T > T_s$ where the replica
calculations yield a replica symmetric result, which in fact
corresponds to ergodicity breaking in an exponentially large
number of states (finite complexity);
\item a low temperature phase, for $T_s > T$, with a
1-step replica symmetry breaking, corresponding to
the predominance of the lowest TAP states, with zero complexity.
\end{itemize}

The relaxation dynamics from a random initial state,
yields equilibrium dynamics in the paramagnetic state for
$T > T_d$, while, for $T_d > T$, the aging phenomena
appears \cite{cukuprl} and the long time limit of the energy per spin
is higher than the equilibrium value.

In this section we review the construction of
the potential function \cite{I}, and
we expose some new results coming from a replica symmetry breaking Ansatz,
which clarify some of the ``mysteries'' left open in \cite{I}.

\subsection{Construction of the potential}

One of the characteristics of spin glasses,
due to their random character, is that
the different equilibrium states
can not be selected by an external field uncorrelated with
the landscape defined by the Hamiltonian. The basic idea underlying
the potential function  is to use an external field pointing in the
direction of a particular equilibrium configuration
\cite{remi,I}. So, if
$\sig_i$ denotes a typical equilibrium configuration at a temperature
$T'$ one can define the partition function
\be
Z[T,\eps,\sig]=\sum_{S} \e^{-\b H[S] +\eps S\cdot \sig}.
\ee
Besides being self-averaging with respect to the distribution of
the quenched Hamiltonian,
the free-energy $\Gamma[T,T',\eps]=-T/N \log Z[T,\eps,\sig]$ is also
independent on the particular
configuration $\sigma$ we choose, and therefore coincide
with its average over the probability distribution $\exp(-\b' H[\sig])/Z[T']$.
We define the ``potential'' as the Legendre transform of $\Gamma$:
\be
V[T,T',q_{12}]=\min_{\eps} \Gamma[T,T',\eps]+\eps q_{12} -F[T].
\ee
 From the Legendre transform we have subtracted $F$ the free-energy at
temperature $T$ in order to have $V[T,T',0]=0$.
Defined in this way, the potential has
the meaning of the free-energy cost to keep a system
at temperature $T$ at fixed overlap $q_{12}$ from a generic configuration
of equilibrium at a different temperature $T'$.

$V$ is self-averaging also with respect to the quenched
disorder distribution, which we denote by an overline.
The basic object  we need to evaluate is then
\be
\ol{{1\ov Z[T']} \sum_{\sig}\e^{-\b'H[\sig]} \log
\l( \sum_{S} \e^{-\b H[S] +\eps S\cdot \sig} \ri)},
\label{bas}
\ee
As it is explained in detail in \cite{I},
in order to perform the averages it is possible to use a double analytic
continuation from integer values of
the parameters $m$ and $n$, used to represent
$\log Z[T,\eps,\sigma]$ as $\lim_{m\to 0}(Z[T,\eps,\sigma]^m-1)/m$, and
${1/ Z[T']}$ as $\lim_{n\to 0}Z^{n-1}$.
There are then
$n$ replicas $\sig_a$ ($a=1,...,n$), and $m$
replicas $S_\a$ ($\a=1,...,m$). The ``external field'' terms is an interaction
term of all the replicas $S_\a$ with one privileged replica, say,
$\sig_1$.
Three overlap matrices turn out to be relevant for the description
of the physics of the model:
$Q^*_{ab}={1\ov N}\sum_i \la \sig^a_i\sig^b_i\ra$ describing the
overlap
statistics of the replicas at equilibrium at temperature $T'$,
$P_{a,\a}={1\ov N}\sum_i \la \sig^a_i S^\a_i\ra$
  describing the overlaps among the replicas at temperature $T'$
and the replicas at temperature $T$, and finally
$Q^{\a\b}={1\ov N}\sum_i \la S^\a_i S^\b_i\ra$
describing the overlaps between replicas at temperature $T$.
As is physically clear, it is found that the
structure of the matrix $Q^*_{ab}$ is not affected at the leading
order by the presence of the replicas $S^\a$. In this paper we will
restrict ourselves to the temperature range
$T\geq T_S$, where $Q^*_{ab}=\delta_{ab}$.
In this regime it is sensible to assume $P_{a,\a}=\delta_{a,1}q_{12}$ for all
$\a$. The structure of the matrix $Q_{\a\b}$ is more subtle. Assuming
a single state picture in ref. \cite{I} the form $Q_{\a\b}=\delta_{\a\b}
+q(1-\delta_{\a\b})$ was taken. But it turns out also to
be necessary to consider
the possibility that ergodicity is broken for the system in a ``field'',
with consequent  replica symmetry breaking in $Q_{\a\b}$. The most
general Ansatz we shall need is the ``one step'' form (see e.g. \cite{mpv}),
characterized by the parameters $(q_0,q_1,x)$. With this Ansatz
it easy to find that the potential
as a function of all the order parameters is:
\bea
V (\q) = -\frac{1}{2\beta}  \left\{
2\beta \beta' f(\q) - \beta^2 \(( (1-x) f(q_1) + x f(q_0) \))
+\frac{x-1}{x}\ln(1-q_1) \right. \\
\left. + \frac{1}{x}
\ln \((1-(1-x) q_1 -x q_0 \))
+\frac{q_0 - \q^2}{1-(1-x) q_1 -x q_0} \right\}
\eea
where $V$ has to be maximized with respect to $q_0$, $q_1$ and
$x$. These saddle point equations read:
\bea
& &\q^2 = q_0 - \beta^2 f'(q_0) (1-(1-x) q_1 -x q_0)^2 , \\
& &\beta^2 (f'(q_1) - f'(q_0))(1-x)= (1-x)\frac{q_1 - q_0}
{(1-q_1)(1-(1-x) q_1 -x q_0)}  , \nonumber \\
& &\beta^2 (f(q_1) - f(q_0)) + \frac{1}{x^2}
\ln \(( \frac{1-q_1}{1-(1-x) q_1 -x q_0} \))
+ \beta^2 \frac{(1-q_1)}{x}f'(q_1)
-\beta^2 \frac{f'(q_0)}{x}(1-(1-x) q_1 -x q_0) = 0.
\label{saddle}
\eea
A numerical resolution allows to construct the curve $V (\q)$
\footnote{In \cite{I}, the form  $Q_{\a\b}=\delta_{\a\b}
+q(1-\delta_{\a\b})$ yielded simplified equations, corresponding
to $q_0=q_1$ in (\ref{saddle}). The resulting potential will be denoted as
``replica symmetric'' potential.}.

In general, this curve can be divided in three regions.
There are a small and a large $q_{12}$ regions
(outside the interval $A-B$ in figure 1) where replica symmetry holds.
In between the symmetry is broken. In the large $q_{12}$ region,
the solution is $q_1=q_0$ testifying ergodicity in a single state.
In the point $B$ a de Almeida Thouless instability develops. The
replica symmetry breaking region is interpreted as usual ergodicity breaking
with dominance of small number of valleys for typical samples.
In the point $A$ one finds $x=1$, and the restoring of replica symmetry
implies in fact a number of valleys  exponentially large
${\cal N}\sim \e^{N\Sigma(q_{12})}$.
In this region (between $\q = 0$ and $A$), $x=1$,
the Edwards-Anderson parameter
inside the valleys is obtained as the value of $q_1$ from the second equation
of (\ref{saddle}) divided by $(1-x)$ in $x=1$,
and is depicted with crosses in figure (\ref{q0q1rsb}).
The complexity $\Sigma(q_{12})$ can be calculated as in the usual case
as ${\partial V\ov \partial x} |_{x=1}$, and is depicted in figure
(\ref{sigma}).
For $q_{12}=0$, where there is no effective constraint, the second
replica is at equilibrium at $T$ and we find
the total complexity at temperature $T$, and the equilibrium
Edwards-Anderson parameter at $T$.

\begin{figure}
\centerline{\hbox{
\epsfig{figure=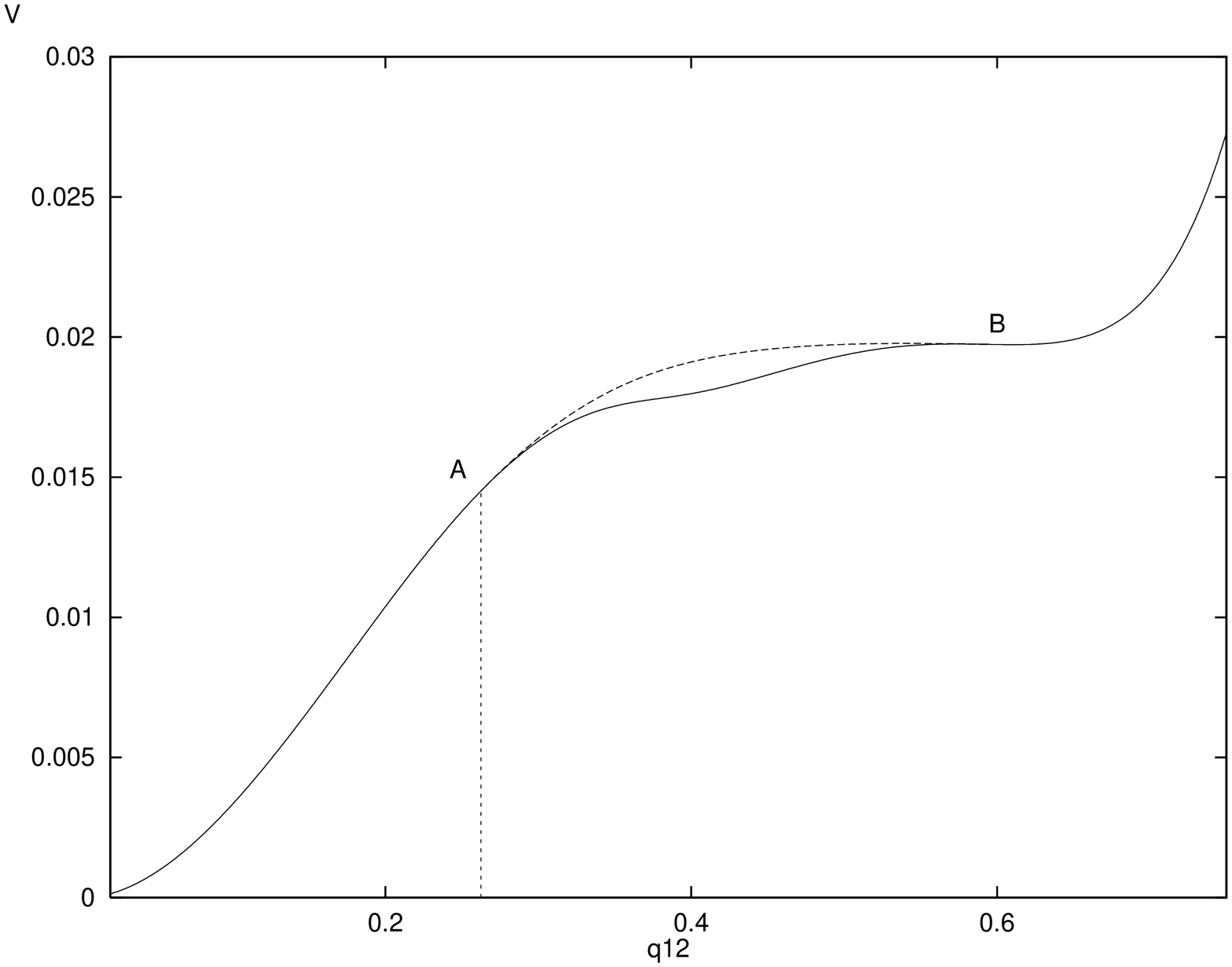,width=5cm,angle=-90}
}}
\caption{Potential $V$ as a function of $\q$, for
the $p=3+p=4$, $\beta=1.25$, $\beta'=1.243$; full line: replica
symmetric solution; dotted line: RSB solution, from A ($x=1$) to
B (where $q_0$ and $q_1$ merge)}
\label{potrsrsb}
\end{figure}

\begin{figure}
\centerline{\hbox{
\epsfig{figure=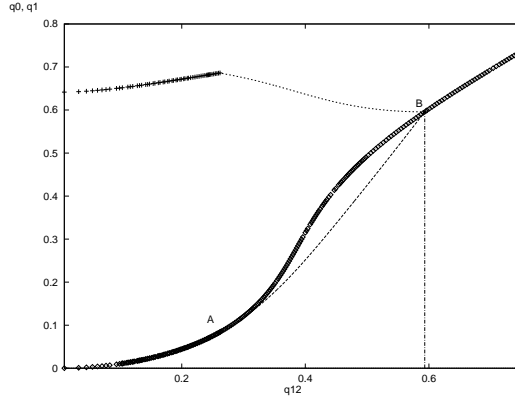,width=5cm,angle=-90}
}}
\caption{$q_0(\q), q_1(\q)$ as a function of $\q$, for
the $p=3+p=4$, $\beta=1.25$, $\beta'=1.243$ (lines); diamonds:
RS solution, given by inserting $q_0=q_1$ in (\ref{saddle});
crosses: continuation of $q_1(\q)$ in the first RS region,
with $\lim_{\q \to 0}q_1(\q) = q_{EA}(\beta)$}
\label{q0q1rsb}
\end{figure}

\begin{figure}
\centerline{\hbox{
\epsfig{figure=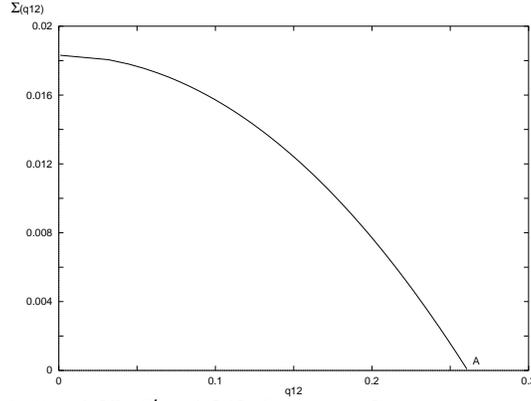,width=5cm,angle=-90}
}}
\caption{$\Sigma (\q)$ for the $p=3+p=4$, $\beta=1.25$, $\beta'=1.243$;
for $\q \to 0$ we recover the complexity at $\beta$. In $A$ the
complexity goes to zero, corresponding to the entrance in
the RSB region of the potential.}
\label{sigma}
\end{figure}

The global situation is displayed for a typical example
in figures (\ref{potrsrsb}) and (\ref{q0q1rsb}).

\subsection{Minima of the potential}

\begin{figure}
\centerline{\hbox{
\epsfig{figure=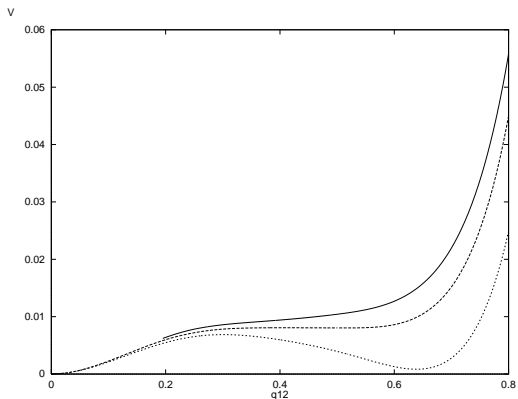,width=5cm,angle=-90}
}}
\caption{Potential $V$ as a function of $\q$, for
the $p=3$ $p$-spin model, for $\beta=\beta'$, from top to bottom
$\beta=1.6$, $\beta=\beta_d \approx 1.633$, $\beta=1.7$;
here $\beta_S=1.706$}
\label{pot}
\end{figure}

The qualitative features of the potential are largely independent of the form
of the function $f$. Let us briefly discuss the case of equal temperatures
$\beta = \beta'$ \cite{I}. The potential has always an absolute
minimum for $\q=0$, corresponding, as previously mentioned,
to the second replica being at
equilibrium. Another minimum appears for a non-zero
value of $\q$ for temperatures below $T_d$ (see figure
(\ref{pot})) (at $T=T'=T_d$, it is a horizontal flex).
This relative minimum
corresponds to having both replicas in the same state, with
$\q = q = q_{EA}$. Since the number of equilibrium states at temperature
$T$ is $\exp(N\Sigma (T))$ by definition of the complexity
(or configurational entropy) $\Sigma$, the free energy cost
of having this situation is
\be
V_{relative \  minimum}(q_{EA})= T \Sigma (T) =
-\frac{\beta}{2}f(q_{EA}) - \frac{q_{EA}}{2\beta} -
\frac{1}{2\beta} \ln(1-q_{EA}).
\ee

For the $p$-spin model, the case of different temperatures $\beta$ and
$\beta'$, has also be treated in some details \cite{I}.
The relative minimum, which still exists for
$0 < T < T_{final}(T')$, and $T' < T_d$,
remains in the replica symmetric region
of the potential, and can be clearly interpreted.
Indeed, the homogeneity of the Hamiltonian allows to write the TAP free
energy in a simple form \cite{kurparvir,I,crisomtap,babumez}:
\be
f_{TAP}(\{m_i \} )=q^{\frac{p}{2}} E^0(\{ \hat{s}_i \} )
 - \frac{T}{2}\ln(1-q) -
\frac{1}{4T}[(p-1)q^p - pq^{p-1} +1] \ ~;
\label{ftap}
\end{equation}
where we have written $m_i = \langle s_i \rangle = \sqrt{q} \hat{s}_i$,
with $\sum_i \hat{s}_i^2 = N$, and the angular energy
(zero-temperature energy) is~:
\be
E^0 (\{ \hat{s}_i \} )
\equiv - \frac{1}{N}
\sum_{1\leq i_1<\cdots<i_p\leq N}\, J_{i_1,\ldots,i_p}\,
       \hat{s}_{i_1}\cdots \hat{s}_{i_p} \ .
\ee

The order in free energy of the solutions of the TAP equations
does not depend on temperature, nor a solution can bifurcate
as the temperature is changed.
All these solutions can be easily
parameterized \cite{kurparvir,I,crisomtap,babumez}
and followed with temperature.

It is then easy to show that the properties of
the primary minimum (value of $q_0=q_1$, and energy)
are precisely the properties of the TAP states
of equilibrium at $T'$ (characterized by a zero
temperature energy $E_0'$) followed at $T$ (parameter $q$ and energy),
with
\be
V_{primary} = T \Sigma(T') + F_{TAP}(T,E_0') -F(T)
\ee.

These situation corresponds therefore to having the second replica
in a TAP state of equilibrium at $T'$ followed at $T$
\footnote{
Let us also note that, for $\beta'=\beta_d$, the minimum becomes
in fact a horizontal flex of the potential, with energy and
parameter $\q$ equal to those obtained in off-equilibrium dynamics.}.
These situation was also ascertained by the study of the dynamics
of a system thermalized at $T'$, and whose temperature was
then changed to $T$ \cite{I,babumez,barrat}: the dynamics obtained
is indeed of equilibrium in these particular TAP states, chosen by
the thermalization at $T'$ and followed when the temperature of the
system is changed.

In the case of $f$ different from a monomial, i.e. of an inhomogeneous
Hamiltonian, many points remained unclear. In particular,
the TAP free energy cannot be parameterized in such a simple form,
and it is not granted that the TAP solutions keep their order in free
energy when the temperature changes.
Moreover, the role of the breaking of replica symmetry was not studied.
In the $p$-spin model, as a consequence of the absence of bifurcation of the
solutions, the minimum of the potential is always in the replica-symmetric
region, and the inclusion of replica symmetry breaking effects does not
affect the discussion of the metastable states, except
for eliminating the spurious secondary minimum found in
\cite{I}, which meaning was not clear.

For an inhomogeneous Hamiltonian,
by studying the potential, including RSB effects, and
the dynamics with thermalized initial conditions,
we will show that the potential still allows to determine
the characteristics of TAP states, as long as the minimum is in the
replica symmetric region. We will associate the entrance of the
minimum in the RSB region of the potential with bifurcations,
and show that, in this case, the dynamics with thermalized
initial conditions gives rise to a particular form of aging.

\section{Potential for an inhomogeneous Hamiltonian}

Let us turn to the detailed study of the potential
in the case of an inhomogeneous Hamiltonian. The numerical
examples will be given for $f(q)=(q^3+q^4)/2$ for simplicity,
but the analysis is independent of this particular form.

We first note that, like for the homogeneous case, a minimum
with $\q \ne 0$ only exists for $T' < T_d$. Moreover, we will limit
us to $T' > T_s$.
For $T=T'$, the primary minimum is still in the replica symmetric
part of the potential.
If $T$ is raised, this minimum stays in the RS region, and disappears
at a certain temperature $T_{final}(T')$, which verifies
$T_{final}(T_d) = T_d$.

As $T$ is lowered however, the endpoint of
the RSB region (where $q_0=q_1 \equiv q$) gets closer to the minimum, and
finally reaches it at $T_{rsb}(T')=\frac{1}{\beta_{rsb}}$ given by
\bea
\beta_{rsb}^2 f''(q) (1-q)^2 = 1 \\
\beta_{rsb}^2 (1-q)^2 f'(q) = q - \tilde{p}^2  \nonumber \\
\beta_{rsb} \beta' f'(\tilde{p}) = \frac{\tilde{p}}{1-q} ,
\eea
where $\tilde{p}$ is the value of $\q$ in the minimum.
For even lower temperatures, the minimum is within the RSB
region.

\begin{figure}
\centerline{\hbox{
\epsfig{figure=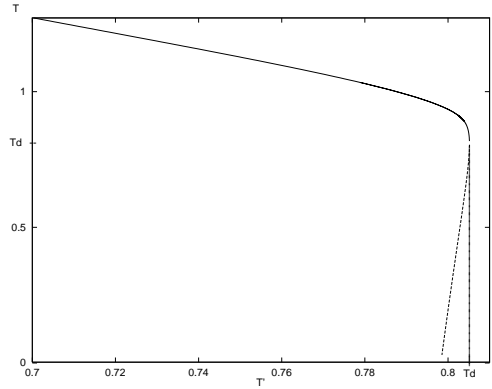,width=5cm,angle=-90}
}}
\caption{$T_{final}(T')$ (full line) and $T_{rsb}(T')$ (dashes)
for $f(q)=1/2(q^3+q^4)$; $T_d \approx 0.805166$. Note that at low enough
temperature, the states never bifurcate. The vertical line corresponds to
$T'=T_d$, i.e. the temperature of appearance of the minimum: along this
line the potential displays an horizontal flex.}
\label{limites}
\end{figure}

$T_{rsb}(T')$ reaches zero for some $T'$ (see figure (\ref{limites}));
for lower $T'$, the minimum is always in the RS region.

\begin{figure}
\centerline{\hbox{
\epsfig{figure=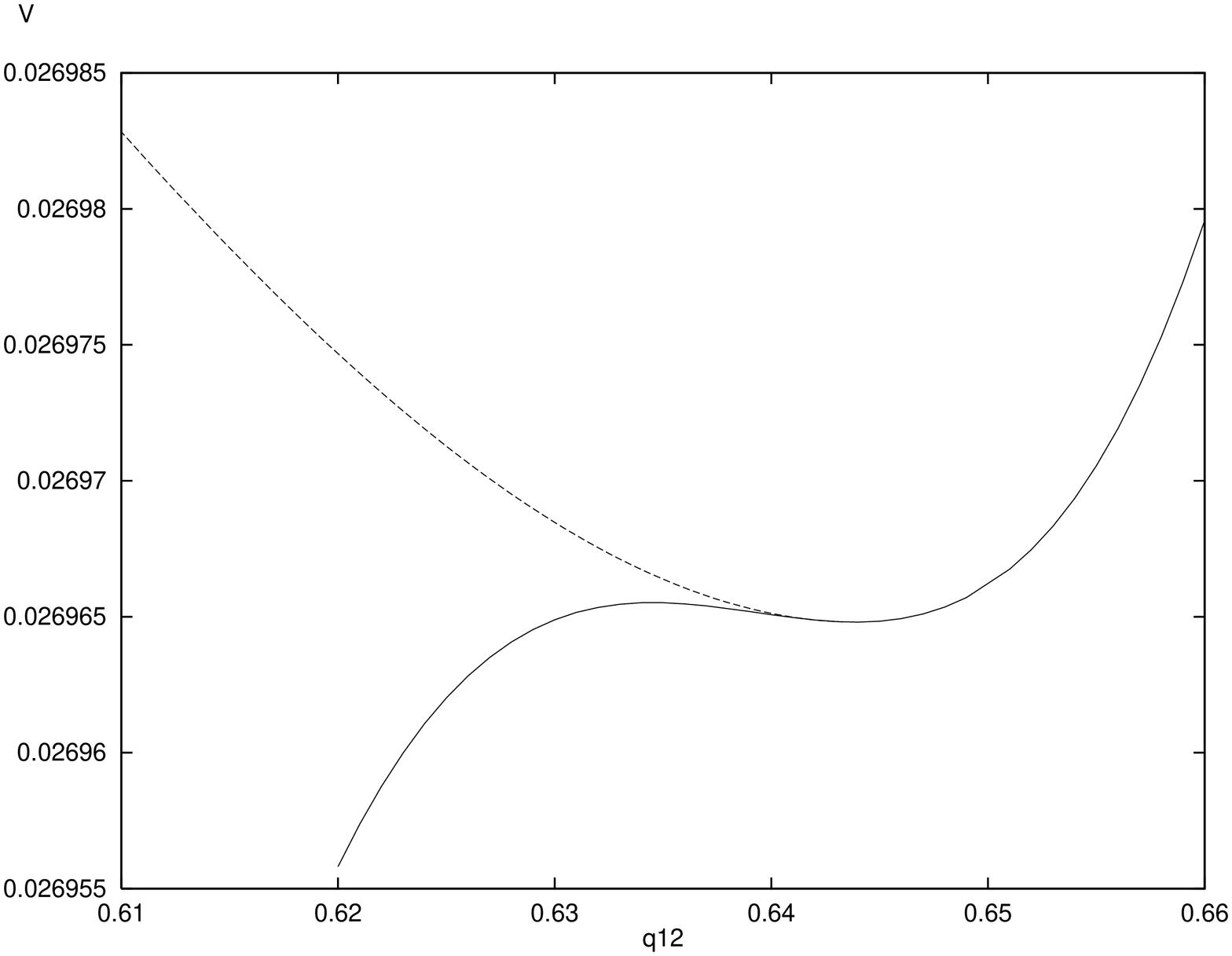,width=5cm,angle=-90}
\epsfig{figure=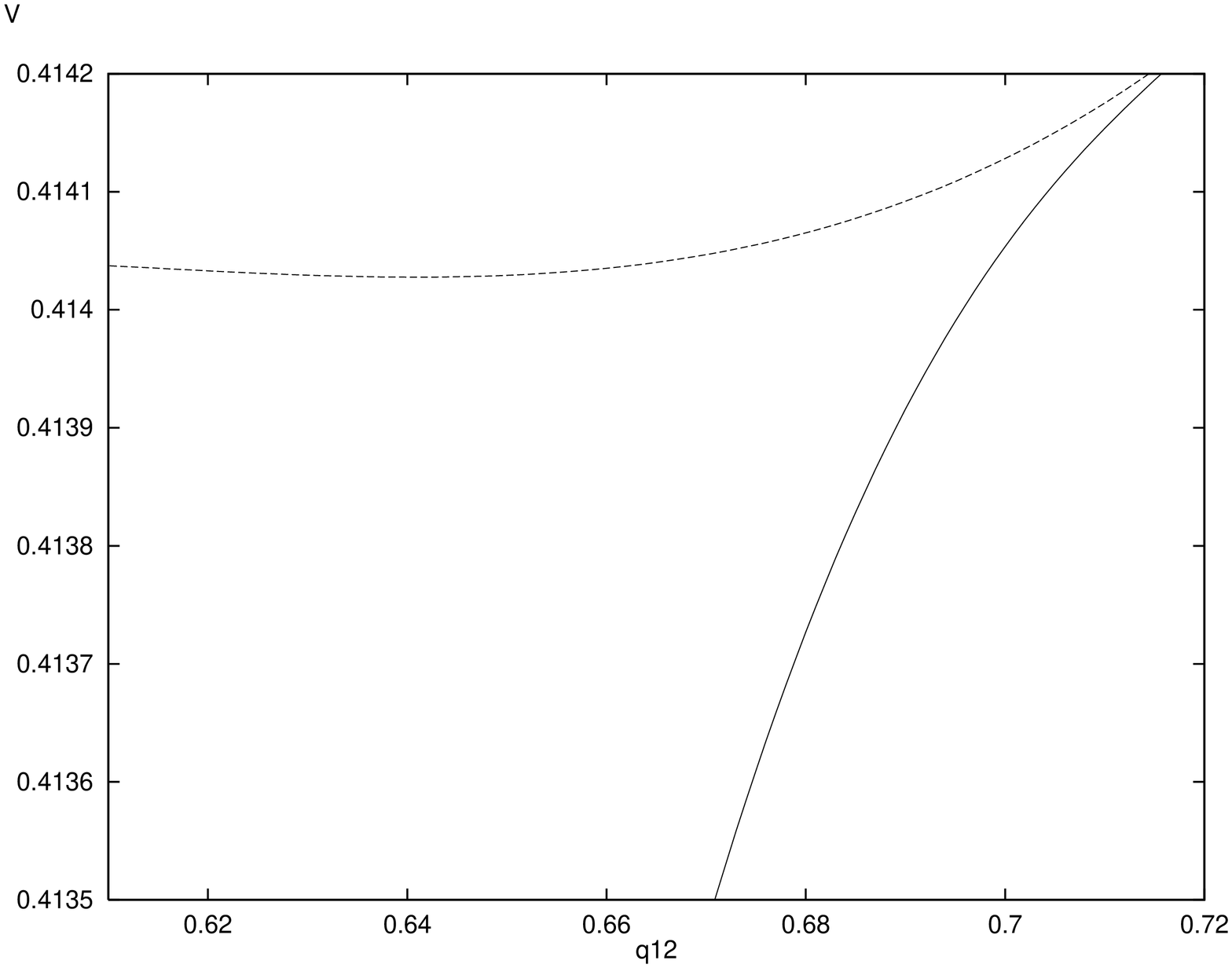,width=5cm,angle=-90}
}}
\caption{Potential for $p=3+p=4$, for $\beta'=1.243$
and $\beta=1.4625$ (left), $\beta=3$ (right);
full lines=RS solution , dashes=RSB solution.
For $\beta=1.4625$ the endpoint of the RSB solution coincides with the
minimum of the potential; for $\beta=3$ the minimum has disappeared
from the RS solution, while it still exists for the RSB curve}
\label{p34}
\end{figure}

An example of the situation $T > T_{rsb}(T')$ is displayed in figure
(\ref{potrsrsb}), while the limiting
case $T=T_{rsb}(T')$ and a case where $T < T_{rsb}(T')$ are shown
in figure (\ref{p34}).

For temperatures $T_{final}(T') > T > T_{rsb}(T')$
the primary minimum can be
interpreted as the state of equilibrium at temperature $T'$ followed down
at temperature $T$.
Indeed, if we consider the TAP states with values of the energy and
of the parameter $q$ equal to those of the primary minimum, and
if we compute their free energy $f_{TAP}$,
we obtain (see appendix):
\be
V_{primary} = f_{TAP} - F_{RS}(T).
\label{vf}
\ee
Following the computation of \cite{riegertap,crisomtap},
it is also possible to obtain the number of TAP solutions with
fixed parameter $q$ and energy $E_{TAP}$,
and, writing it in the form
\be
\exp(N S(q,E_{TAP},T))
\ee
we have checked numerically the identity:
\be
S(q^{pr},E_{primary},T) = \Sigma(T').
\ee
Therefore, the number of equilibrium TAP solutions at $T'$
($\exp(N \Sigma (T'))$) is equal to the number of TAP solutions at $T$
with the energy and the parameter $q$ of the primary minimum.
This fact, together with (\ref{vf}), shows that the state
of equilibrium at $T'$ has been followed at $T$,
and is a stable state with free energy cost
\be
V_{primary} = T\Sigma(T') + F_{TAP}(E_{primary},q^{pr},\beta) - F_{RS}(T).
\ee
(where the total, i.e. with the complexity term,
free energy of the TAP states is
$F_{TAP}(E_{TAP},q,\beta)=f_{TAP}(E_{TAP},q,\beta) - T S(q,E_{TAP},T)$).
%(calcul de S en appendice ??)
We will show in next section how
these states can be followed dynamically, by choosing appropriate
initial conditions.

For $T < T_{rsb}(T')$ on the contrary, the primary minimum
is in the region of the potential which displays replica
symmetry breaking
\footnote{Note that $T_{rsb}(T_d)=T_d$, and that, for $T'=T_d$, $T<T_d$,
the minimum
is in fact an horizontal flex, like for the homogeneous case, except that
it lies within the RSB region of the potential. Besides, the energy
in this point is equal to the dynamical energy at $T$.}.
The obvious interpretation for this is that at
$T_{rsb}(T')$, the metastable states multifurcate, according to
the usual pattern known from the physics of the Sherrington-Kirkpatrick
model.

Let us now address the problem of level crossing: the $p$-spin model
seems very particular, in that the order in free energy of
the TAP states does not depend on temperature. For $T < T_s$,
the statics are given by the lowest TAP states, therefore there
are high correlations between equilibrium states at different
temperatures. On the other hand, for temperatures between
$T_s$ and $T_d$, equilibrium measures at different temperatures
are given by different bunches of TAP states; therefore the
overlap between equilibrium states at different temperatures
is zero, but the TAP states can be followed at other temperatures,
and their order in free energy (without the complexity term)
remains the same.

For the case of an inhomogeneous Hamiltonian, we also show
that, as long as we consider TAP states giving the equilibrium
measure at temperatures higher than $T_s$, we have
no crossing in the free energies $f_{TAP}$: indeed, if we
note $f(T,T')$ the free energy of one TAP state of equilibrium at $T'$,
followed at $T$, we have
\be
f(T,T') = V_{primary}(T,T') + F(T),
\ee
and thus we obtain
\be
\frac{\partial }{\partial T'} f(T,T') = \frac{f(\q^{pr})}{T'^2}.
\ee
This quantity is always positive, so, if we have two temperatures
$T_d > T'_1 > T'_2 > T_s$, at any temperature $T$ for which we can
follow the states giving equilibrium at $T'_1$ and $T'_2$, the order
\be
f(T,T'_1) > f(T,T'_2)
\ee
is conserved.
Of course, this is not the case if we consider the full free
energy, with the complexity term, i.e.
$F(T,T')=f(T,T') - T \Sigma(T')$. We then have that each
curve $F(T,T')$ as a function of $T$ is tangent to the curve
$F_{RS}(T)$ at the point $(T',F(T',T')=F_{RS}(T'))$.

Hence, this global situation, with the replica symmetric
free energy as the envelope of the curves giving the
total TAP free energies, whereas the curves giving the
TAP free energies without the complexity term do not
cross, seems very generic between the static and dynamic
transitions.

\begin{figure}
\centerline{\hbox{
\epsfig{figure=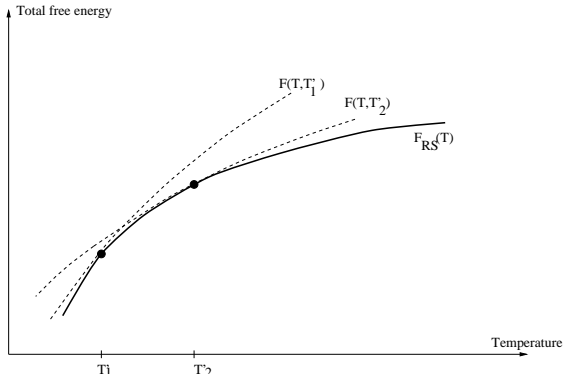,width=5cm,angle=-90}
}}
\caption{Total free energy}
\label{freeenergy}
\end{figure}

Considering  the case of $T'=T_s$, the lowest $T'$ for
which we are allowed to use
the simple Ansatz $Q^*_{ab}=\delta_{ab}$, we find a different situation.
In fact, we find that if $T<T_s$ the value of
the potential in the primary minimum is slightly higher
than 0, with $q_{EA}(T) = q_1(\q \to 0) > q^{pr} > \q^{pr}$.
This is in contrast with the case of the $p$-spin model, for which
the states of equilibrium at
$T_s$, followed at $T$, are still of equilibrium at
$T$: these are the lowest TAP states, and they dominate the
equilibrium measure for temperatures ranging from $0$ to $T_s$.
In this case we obtain $V_{primary}=V(0)$, and $q^{pr}=q_1(\q \to 0)$.
Here on the contrary, the difference between the quantities at
$\q = 0$ and at the primary minimum show that the states of
equilibrium at $T_s$ are no more of equilibrium at $T < T_s$.
Therefore, there is presence of chaos in temperature.
For a detailed study of the $T' < T_s$ region, we would however
need to take into account the RSB effects on the first replica,
which would yield another form for the potential, and we will not do it here.

\section{Dynamics}

We now address the problem of the dynamics of the system
at $T$, starting from thermalized initial
conditions at $T'$.
In the case of the $p$-spin model, it was shown that such a
procedure allows to reach dynamically the states
described by the minimum of the potential, i.e. to
follow dynamically the TAP states.
As usual, we study the Langevin relaxation dynamics of the model,
given by
\be
\frac{ds_i(t)}{dt} = -\frac{\partial H}{\partial s_i} - \mu(t) s_i(t)
 + \eta_i(t),
\label{langepspin}
\ee
where the $\eta_i$ are Gaussian thermal noises with
$\langle \eta_i(t) \eta_j(t') \rangle = 2T\delta_{ij}
\delta(t-t') $, and $\mu(t)$ has to be computed self consistently
in order to implement the spherical constraint
$\sum_{i} s_i^2 = N$.

In the infinite $N$ limit, we can obtain
the dynamical equations for the correlation and response functions
($C(t,t') = \frac{1}{N} \sum_i \overline{\langle s_i(t)s_i(t') \rangle}$,
$r(t,t') = \frac{1}{N} \sum_i
\overline{\langle \frac{\partial s_i(t)}{\partial \eta_i(t')} \rangle}$)
\cite{I,houghton}, that for $t>t'$ read :

\bea
{\partial r(t,t') \over \partial t}&=& -\mu(t)  r(t,t')
+ \int_{t'}^t ds  \ f''(C(t,s))r(t,s)r(s,t')
\nonumber \\
{\partial C(t,t') \over \partial t} &=&-\mu(t)C(t,t')
+ \int_0^{t'} ds \  f'(C(t,s))r(t',s) \nonumber \\
&+& \int_0^t ds  \ f''(C(t,s))r(t,s)C(s,t')
+ \frac{1}{T'} f'(C(t,0))\ C(t',0) .
\label{eqTT}
\eea
complemented by the equation that enforces the spherical condition
\bea
\mu(t) &=& \int_0^t ds\  f'(C(t,s)) r(t,s)
+ \int_0^t ds\  f''(C(t,s)) r(t,s) C(s,t) \nonumber \\
&+& T + \frac{1}{T'} f'(C(t,0))C(t,0)
\eea

In \cite{I}, it was noted that a numerical integration of
(\ref{eqTT})  for a particular choice of the temperatures,
after some transient led to equilibrium  with time
translation invariance (TTI) and validity of the fluctuation dissipation
theorem (FDT). However, no systematic study was undertaken.

As long as the primary minimum of the potential is in the replica
symmetric region it is reasonable to take as an
Ansatz that indeed an equilibrium regime is
reached after a short transient.
We therefore deal with the functions $C_{as}(\tau)$,
$r_{as}(\tau)$ related by FDT, with the introduction of
the limiting quantities $\tilde{p}$ and $q$:
\bea
C(t,t') = C_{as}(t-t') ; \ r(t,t') = r_{as}(t-t') ;
r_{as}(\tau) = -\beta \frac{\partial}{\partial \tau}C_{as}(\tau)
\nonumber \\
\lim_{t \to \infty} C(t,0) = \tilde{p}
 ; \lim_{\tau \to \infty} C_{as}(\tau) = q
\eea

This Ansatz yields the same equations for $\tilde{p}$
and $q$  as the ones for $q_{12}$ and $q$ (\ref{eqrs})
specifying the extremum of the potential in the RS region
\cite{I}.
Besides, it coincides very well with the results of a
numerical integration of equations (\ref{eqTT}).
We can therefore conclude that the dynamics takes place in
a TAP state, of equilibrium at $T'$, in which the system
was put by thermalization at $T'$,
followed dynamically at the new temperature $T$. This behavior
is exactly the same as for the $p$-spin model \cite{babumez}.

For $T < T_{rsb}(T')$ another Ansatz has to be chosen.
In particular, since for low enough $T$ the minimum disappears
from the RS potential, the dynamical equations for
$\tilde{p}$ and $q$ have no more solutions.
We therefore propose an Ansatz
similar to the one used in the aging dynamics of such models
\cite{cukuprl}, except that the motion will
be confined in the vicinity of the initial state. We assume then that:
\begin{itemize}
\item for finite time separations $\tau=t-t'$, with $\tau /t$
going to zero, the equilibrium properties are valid, which means that
we deal with the functions
$C_{FDT}(\tau)$ and $r_{FDT}(\tau)$, related by FDT. We note
$\lim_{\tau \to \infty} C_{FDT}(\tau)= q_1$.

\item an aging regime is present:
for $t$ and $t'$ going to infinity, without
$(t-t')/t \to 0$, time translation invariance is violated, and
the FDT is replaced by the quasi-FDT
\be
x \frac{\partial C(t,t')}{\partial t'} = T r(t,t'),
\ee
with constant $x \ne 1$. In this regime, we have the limits
$\lim_{t'/t \to 1}C(t,t') = q_1$,
$\lim_{t'/t \to 0}C(t,t') = q_0$.

\item we have moreover to introduce the quantity
$\lim_{t \to \infty} C(t,0) = \tilde{p}$, which tells how much the system
remembers its initial conditions.
\end{itemize}
As it happens in the random initial condition case, the
parameters $q_1,q_0,\tilde{p},x$  can be determined from the asymptotic
analysis of equation (\ref{eqTT}) without fully solving the dynamics.
The hypothesis of existence of an aging regime, and the continuity
of the response function implies the equation
\be
\b^2f''(q_1)(1-q_1)^2 =1
\ee
which coincides with the ``marginal stability condition''
of the statics \cite{margin,cukuprl}.
The other three equations
\bea
\frac{q_1}{\beta(1-q_1)} &=& \beta f'(q_1) (1-q_1) +
\beta x (q_1 f'(q_1) - q_0 f'(q_0)) + \beta' \tilde{p} f'(\tilde{p})
 \nonumber \\
\frac{\tilde{p}}{\beta(1-q_1)} &=& \beta \tilde{p} x (f'(q_1) - f'(q_0))
+ \beta' f'(\tilde{p})
 \nonumber \\
\frac{q_0}{\beta(1-q_1)} &=& \beta f'(q_0) (1-q_1) +
\beta q_0 x (f'(q_1) - f'(q_0)) + \beta' \tilde{p} f'(\tilde{p}),
\eea
can be shown to be equivalent the vanishing of the derivatives
of the potential function (\ref{pot}) with respect to $q_1$,
$q_0$ and $q_{12}$.
In terms of these parameters
the asymptotic energy is given by:
\be
E = -\beta' f(\tilde{p}) -\beta (f(1)-f(q_1)) -\beta x (f(q_1)-f(q_0)).
\label{energy}
\ee
The usual aging behavior with $q_0 = \tilde{p} =0$
is of course solution of these equations. This corresponds
to forgetting the initial conditions, and happens when
$T'$ is higher than the dynamical transition temperature.
However, for $T' < T_d$, this solution, besides of contrasting
with the statical picture of the model,
 would be internally contradictory:
indeed, at $T_{rsb}$ the energy in the primary minimum is lower
than the dynamical energy at the same temperature. Therefore, such
a solution, which would yield an asymptotic energy equal
to the dynamical one, would lead to a higher energy for
a lower temperature!
Finally, the numerical integration of the dynamical
equations  shows
that the behavior of the dynamical quantities is very
different from the case of infinite $T'$, and that
$C(t,0)$ does not seem to decrease to zero.
This facts lead to the conclusion that we must prefer the solution with
non-zero $q_0$ and $\tilde{p}$.
The aging takes therefore place in a restricted
phase space region.
However, for $T< T_{rsb}(T')$ the dynamic internal energy
is higher then the static one, similarly to what happens starting
from random initial conditions.

Let us also note that the asymptotic energy (\ref{energy}) in the
case of thermalized initial conditions is lower
than the dynamical energy after a quench, showing
that this procedure allows to reach states with lower energies.
An immediate consequence is the importance of the way in which
the final temperature is reached.

\section{A comparison with real glasses}
\subsection{General considerations}

In the studied mean-field models,
we have found that, below the dynamical transition $T_{D}$, we
could define a whole spectrum of internal energies for the system
at temperature $T$, depending on the way the system has been put
at its final temperature:
\begin{itemize}
\item the equilibrium energy $E_{eq}(T)$, which is done by
the usual Boltzmann Gibbs formula;
\item the dynamical energy, corresponding to
the energy of a system which is quenched to the final temperature from a
temperature higher than $T_{D}$;
\item the energies $E(T',T)$, obtained for a system at equilibrium at $T'$
and then put at $T$. Depending on $T'$ and $T$, the system can be
at equilibrium or exhibit aging dynamics.
\end{itemize}

These energies can be consistently computed using the explicit form of
the dynamics. It is also possible to compute them by using the appropriate
statistical prescription which does not make explicite reference
to the dynamics.

At this point the reader may ask how much all these findings are relevant for
the real world.
Metastable states with {\em infinite} life do not exist in short range finite
dimensional models and
their presence in mean field models is a clear artefact of the approximation.
The would be infinite
life metastable states of the mean field theory do decay through some activated
processes (whose detailed properties have not yet been fully clarified).
If the mean field picture is relevant for the
real word the time scale of the activated processes should be large enough that
there is a time window in which the behavior predicted by
mean field theory can be observed.

Given our lack of command on the activated processes, we cannot treat
this question analytically and
we have to resort to numerical simulations. We will consider a simple system,
one of the prototypes of glass forming systems, known to have a glass
transition at a given temperature $T_G$.

We will see that we can also define various energies:
\begin{itemize}
\item the equilibrium energy $E_{eq}(T)$;
\item the slow cooling energy $E_{S}(T)$, which is obtained by the limit to
infinite cooling time of
the energy of a system which starts at temperatures greater than the dynamical
transition;
\item the fast cooling energy $E_{F}(T)$, which is obtained by the limit to
infinite cooling rate of
the energy of a system which is quenched to the final temperature from a
temperature higher than $T_{D}$.
\end{itemize}
Between $E_{S}(T)$ and $E_{F}(T)$, various cooling rates will
yield various asymptotic energies.

We will see that if we fast cool the system  to a
temperature near of below $T_G$
(we have investigated up to temperatures equal to $.25 T_G$) the energy as
function of the time may be represented by the following form
\be
E_F(t)= E_F+A t^{-\mu} +O(t^{-2\mu}),\label{TIME_FAST}
\ee
where the exponent $\mu$ is in the range $.5-.7$ and weakly depends on the
temperature. The previous
formula well represent the data for time in the window $10^2 -10^5$ time units
(i.e. one Monte Carlo sweep).

In the similar way we can represent the data for the slow cooled energy as
function of time with a
similar form in the same time window:
\be
E_S(t)= E_S+A t^{-\mu} +O(t^{-2\mu}),\label{TIME_SLOW}
\ee
where the exponent $\mu$ is compatible to be equal to the one used
in equation (\ref{TIME_FAST}).

The two functions $E_F$ and $E_S$ are different one from the other below $T_G$
and their difference vanishes when we approach $T_G$.
It is clear that the two previous formulae can be valid only in a
limited time window;
asymptotically the two energies $ E_F(t) $ and $E_S(t)$ must go to the same
limit, (i.e. the equilibrium value of the energy). This is likely
to happens on a much longer scale. Here we want to
stress the presence of a time window in which the prediction
of a theory based on the existence of metastable states can be tested.

Before showing the results of the numerical simulations, we will give some
details of the model we consider.

\subsection{The Hamiltonian}
The model we consider is
the following.  We have taken a mixture of soft particles of different sizes.
Half of the particles
are of type $A$, half of type $B$ and the interaction among the particle is
given by the Hamiltonian:
\begin{equation}
H=\sum_{{i<k}} \left( \frac{\si(i)+\si(k)}{|{\bf x}_{i}-{\bf
x}_{k}|}\right)^{12},\label{HAMI}
\label{HAMILTONIAN}
\end{equation}
where the radius ($\si$) depends on the type of particles. This model has been
carefully studied in the past
\cite{HANSEN1,HANSEN2,HANSEN3,LAPA,PAAGE1,PAAGE2}.  It is known that a
choice of the radius such that $\si_{B}/\si_{A}=1.2$ strongly inhibits
crystallisation and that the
systems goes into a glassy phase when it is cooled. Using the same
conventions as the previous investigators we consider particles of
average diameter $1$, more precisely we set
\begin{equation}
{\si_{A}^{3}+ 2 (\si_{A}+\si_{B})^{3}+\si_{B}^{3}\over 4}=1.
\label{RAGGI}
\end{equation}

Due to the simple scaling behaviour of the potential, the thermodynamic
quantities depend only on
the quantity $T^{\frac{1}{4}}/ \rho$, $T$ and $\rho$
 being respectively the temperature
and the density. For definiteness we have taken $\rho=1$.
It is usual to introduce the quantity $\Gamma^4 \equiv
\beta$. The glass transition is known to happen around $\Gamma=1.45$
(i.e.  for $T \approx 0.226$) \cite{HANSEN2}.

\subsection{Numerical results}

Our simulation are done using a Monte Carlo algorithm, which is more easy to
deal with than
molecular dynamics, if we change the temperature in an abrupt way.  Each
particle is shifted by a
random amount at each step, and the size of the shift is fixed by the condition
that the average
acceptance rate of the proposal change is about .4. Particles are placed in a
cubic box with
periodic boundary conditions.  In our simulations we have considered a
relatively small number of particles i.e. $N=66$.
Note that for all the simulations, the system is always out
of equilibrium and exhibits aging: the ergodic time is far beyond
reach \cite{LAPA,PAAGE1,giorgio100}.

\begin{figure}
\centerline{\hbox{
\psfig{figure=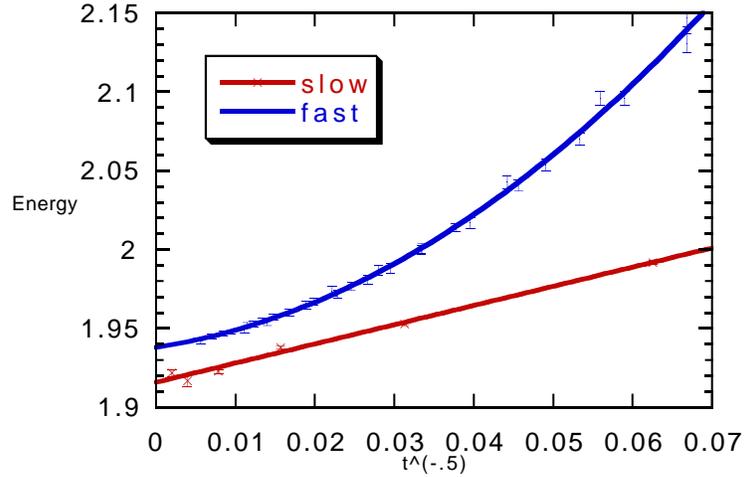,width=10cm}
}}
\caption{``fast'' curve: energy as a function of $t^{-.5}$ where $t$ is
the time spent at $\Ga = 1.8$ after a rapid quench;
``slow'' curve: energy reached at the end of the time spent
at $\Ga = 1.8$, as a function of of $t^{-.5}$, where $t$ is
the time spent at each temperature during the gradual quench
process. We see that slower cooling yields lower energies. The continuation
to $t^{-.5} \to 0$ corresponds to an infinitely slow cooling.}
\label{SIM_1}
\end{figure}

We start by placing the particles at random and we quench the
system by putting it at its final
temperature (i.e. infinite cooling rate). The typical value of the energy
density of the initial
configuration is very high ($O(10^5)$) due to the singular form of the
potential and it takes a few iterations to arrive
to a more reasonable value.
We show the data as function of the Monte Carlo time
$t$ in figure (\ref{SIM_1}) for $\Ga=1.8$ ($T \approx 0.095$).

\begin{figure}
\centerline{\hbox{
\epsfig{figure=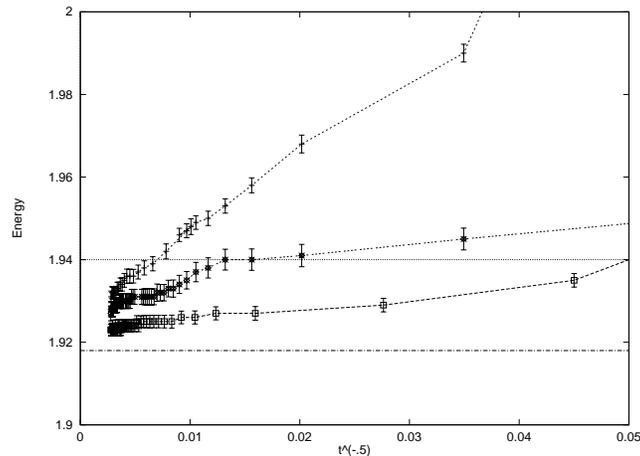,width=6cm,angle=-90}
}}
\caption{Evolution of the energy at $\Ga=1.8$ ($T \approx 0.095$),
in function of time, for various cooling rates; the horizontal lines
correspond to $E_F$ and $E_S$. The lower curves correspond to slower
coolings.}
\label{coolingrates}
\end{figure}

\begin{figure}
\centerline{\hbox{
\epsfig{figure=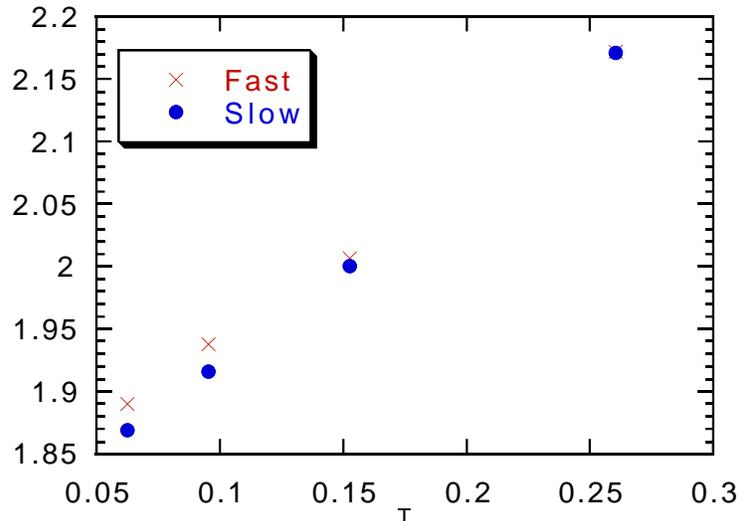,width=10cm}
}}
\caption{Extrapolations of the energies $E_F$ and $E_S$ at large times,
i.e. asymptotic energies after a quench or after an infinitely slow cooling,
for various temperatures.}
\label{SIM_2}
\end{figure}

In the slow cooling approach we also start by placing the particles
at random at the beginning. We divide the
cooling time in $5$ equal intervals: in the first interval we
have $\Ga=1$, in the second interval
$\Ga=1.2$,... and the fifth interval $\Ga=1.8$.
The data are taken for each temperature only in the
second half of the corresponding interval. The  results
as function of the time spent at each temperature, i.e. of the
inverse of the cooling rate, are shown
in figure (\ref{SIM_1}) for $\Ga=1.8$.
We clearly see that the two curves $E_F$ and $E_S$
definitely extrapolate to a different value.
The extrapolated values of the energy as function of the
temperature can be seen in figure (\ref{SIM_2})
using the fast and the slow cooling method in the region $\Ga>1.4$
($T < 0.26$). The data are not shown at higher
temperatures, because the two methods give the same result.
Other procedures to investigate the dependence of the cooling
rate involve a similar cooling from  $\Ga=1$ to
$\Ga=1.8$ in a total time $t_{cooling}$, with times
$t_{cooling}/4$ spent at $\Ga = 1.,\ 1.2,\ 1.4,\ 1.6$,
and then the study of the evolution of the energy at $\Ga=1.8$.
The long time limit of the energy lies then between
$E_F$ (for $t_{cooling} \to 0$) and $E_S$ (for
$t_{cooling} \to \infty$).
We show in figure (\ref{coolingrates}) the evolution
of the energy at  $\Ga=1.8$ for various cooling rates.
The effects are quite small, so
it is necessary to compare reasonably different rates. For the available
times, the energy of the system depends on the cooling it has followed:
the energy is lower for slower cooling procedures.

Moreover, it is worth noting that the value of
$\mu$ is roughly speaking independent from the temperature
\cite{giorgio100}. This phenomenon happens in the only model
of mean field theory where the exponent has been computed
\cite{fraparmar} and this is a strong indication that the
approach to equilibrium in this region is not dominated by activated processes,
but more (roughly speaking) by entropic barriers: the barriers
between metastable states could include both energetic and entropic
effects \cite{BG}.
Moreover, this shows the possible relevance of the scenario detailed
in the preceeding paragraph (aging similar to usual, but at lower
energies), and of some intuitive mean-field
scenarios \cite{LaKu}.

It would be interesting to be able to simulate the thermalization
at a certain temperature, followed by a quench at a lower
temperature, like in mean-field models. Unfortunately, the
available time window do not allow us to reach thermalization
at temperatures lower than the dynamic transition.

Another possibility would be to cool very slowly the system to
a certain value $T'$, such that its energy is lower than
$E_S(T)$ for a certain $T$ ($T > T'$), and then to bring the
system back to $T$, to see whether the obtained energy is still
lower than $E_S(T)$.
Such investigations are however beyond the scope of the
present short study.

\section{Summary and Conclusions}

In this paper we have investigated the behavior in temperature of the
metastable states of long range spin-glasses with first
order freezing transition. We have shown that the metastable states can be
followed up and down in temperature, from the temperature
where they are dominating the partition function.
Going up in temperature, one finds some temperature where
the states disappear, merging with some maxima.
Going down in temperature, the state do never disappear, although
in some range of $T'$ multifurcation is found.
We also studied the dynamics at temperature $T<T_d$,
following a quench from equilibrium at temperature $T'$.
If $T'>T_d$ we find no difference
from  the usual aging behavior \cite{cukuprl} that follows a quench
from infinite temperature. For $T'<T_d$ we have found two possibilities.
If $T>T_{rsb}(T')$ the original valley  has  ``deformed''
but not bifurcated and the system is able to equilibrate inside it.
In the complementary interval $T<T_{rsb}(T')$ the landscape has changed
drastically as the original valley has bifurcated. The system is
then unable to thermalize and falls in an aging regime, while remaining
confined in the vicinity of the initial data.
Besides, this dynamical study shows that the aging after a slow
quench (in the mean-field case, the case of thermalized initial
conditions at $T'$ can be thought of as a situation after an
infinitely slow quench)
allows to reach a situation where the behavior is qualitatively
similar to the one after a rapid quench (i.e. aging
corresponding to a slow touring of the phase space), but
within a phase space region with lower energies.
Therefore, at a given temperature $T$, the possibilities are not only
of aging at a relatively high energy, after a sudden quench, or
of equilibrium dynamics after an infinitely slow quench, but also
of aging at intermediate energies, depending on the route from
high temperature to $T$.

In the last section, we tried to emphasize the possible
relevance of such mean-field scenarios for finite
dimensions, where it has been advocated that metastable states
may still exist, but with a finite lifetime: coming from a high
temperature phase, the system may be able to find these states in
a finite time, and the resulting aging behavior when decreasing
the temperature could be a mixture of jumps between states
and periods of wandering when states bifurcate.

Indeed, the numerical study of section V shows indications that,
at least in the explored time window, for a soft sphere model
of glass exhibiting aging, the dynamics is not
dominated by activated processes. Depending on the cooling rate
from the high temperature phase, various energies can be
reached.
Since the system is finite, it should however reach equilibrium
in a finite time (the energy should reach the equilibrium energy,
whatever may be the route to the final temperature) but, these
simulations show that, even for a relatively small system, this
finite time is very large, and therefore that mean-field
conclusions can be of importance in the real world.

\acknowledgments

It is a pleasure to thank A. Cavagna, I. Giardina and M. Virasoro
for interesting discussions.

\appendix

\section*{}

We consider the case when the primary minimum of the potential
is in the RS region: $q_0=q_1 \equiv q$. Then,
for fixed $T'$, we compute the value of $\q$ and $q$ for this
minimum, $\q^{pr}$ and $q^{pr}$.
The saddle point equations for $q_0,\ q_1,\ x$ reduce to
\be
\beta^2 f'(q^{pr}) = \frac{q^{pr}- {\q^{pr}}^2}{(1-q^{pr})^2}  \\
\ee
and the equation $\frac{\partial V}{\partial \q}= 0$ is
\be
\beta \beta' f'(\q^{pr}) = \frac{\q^{pr}}{1-q^{pr}}.
\label{eqrs}
\ee
The value of the potential is
\be
V_{primary} = -\beta' f(\q^{pr}) + \frac{\beta}{2}f(q^{pr})
-\frac{\beta}{2}(1-q^{pr})f'(q^{pr})
-\frac{1}{2\beta} \ln(1-q^{pr}).
\ee

The energy of the second replica, in this minimum, is
\be
E_{primary} = \frac{\partial}{\partial \beta}
(\beta V + \beta F(T))
\ee
which yields
\be
E_{primary} = -\beta' f(\q^{pr}) + \beta f(q^{pr}) -\beta f(1)
\ee

On the other hand, we can write the TAP free energy as:
\be
f_{TAP}(H,q,\beta) = H - \frac{1}{2 \beta}\ln(1-q) -
\frac{\beta}{2}(f(1) - f(q) - (1-q)f'(q)),
\ee
where $q=\frac{1}{N}\sum_i m_i^2$, and $H$ is the value taken by the
Hamiltonian $H(\{m_i\})$,
so the energy of a TAP state ${m_i}$ is
\be
E_{TAP} = \frac{\partial \beta f}{\partial \beta} =
 H -\beta (f(1) - f(q) - (1-q)f'(q))
\ee

Then, taking
\be
H_{pr}=E_{primary} + \beta (f(1) - f(q^{pr}) - (1-q^{pr})f'(q^{pr}),
\ee
we obtain immediately that
\be
V_{primary} = f_{TAP}(H_{pr},q^{pr},\beta) - F_{RS}(T).
\ee
This means that $V_{primary}$ is the free energy cost of having the
second replica in a TAP state with parameter $q^{pr}$ and energy
$E_{primary}$ at inverse temperature $\beta$.

\end{document}